\documentclass[12pt]{article}

\usepackage{geometry}
\usepackage{lscape}
\usepackage{array}

\usepackage{graphicx}
\usepackage{amsmath}
\usepackage{bbm}

\usepackage{cite}
\usepackage[sort&compress,numbers]{natbib}
\usepackage{amsfonts}
\usepackage{amssymb}
\usepackage{amsmath}
\usepackage{latexsym}
\usepackage{xcolor}
\usepackage{url}
\usepackage{ntheorem}
\usepackage{braket}
\usepackage[normalem]{ulem}
\usepackage{enumerate}

\usepackage[colorlinks,citecolor=blue,urlcolor=blue,bookmarks=false,hypertexnames=true]{hyperref}

\newcommand{\ptcheck}[1]{\ptc{checked on #1}}

\newcommand{\mcO}{\mycal O}

\newcommand{\redM}{{M}}

\DeclareFontFamily{OT1}{rsfs}{}
\DeclareFontShape{OT1}{rsfs}{CGNPm}{n}{ <-7> rsfs5 <7-10> rsfs7 <10-> rsfs10}{}
\DeclareMathAlphabet{\mycal}{OT1}{rsfs}{CGNPm}{n}

\newtheorem{theorem}{\sc  Theorem\rm}[section]

\newtheorem{corollary}[theorem]{\sc  Corollary\rm}

\newtheorem{lemma}[theorem]{\sc Lemma\rm}

\newtheorem{proposition}[theorem]{\sc Proposition\rm}

\newtheorem{remark}[theorem]{\sc Remark\rm}
\newtheorem{Remark}[theorem]{\sc Remark\rm}

{\catcode `\@=11 \global\let\AddToReset=\@addtoreset}
\AddToReset{equation}{section}

{\catcode `\@=11 \global\let\AddToReset=\@addtoreset}
\AddToReset{figure}{section}

{\catcode `\@=11 \global\let\AddToReset=\@addtoreset}
\AddToReset{table}{section}

\newcommand{\R}{\mathbb R}

\newcommand{\hyp}{ \mycal S}

\newcommand{\mcH}{\mycal H}

\newcommand{\mcD}{{\mycal D}}
\newcommand{\mcN}{{\mycal N}}

\newcommand{\bea}{\begin{eqnarray}}
\newcommand{\beaa}{\begin{eqnarray*}}
\newcommand{\bean}{\begin{eqnarray}\nonumber}

\newcommand{\bel}[1]{\begin{equation}\label{#1}}
\newcommand{\beal}[1]{\begin{eqnarray}\label{#1}}
\newcommand{\beadl}[1]{\begin{deqarr}\label{#1}}
\newcommand{\eeadl}[1]{\arrlabel{#1}\end{deqarr}}
\newcommand{\eeal}[1]{\label{#1}\end{eqnarray}}
\newcommand{\eead}[1]{\end{deqarr}}
\newcommand{\eea}{\end{eqnarray}}
\newcommand{\eeaa}{\end{eqnarray*}}

\newcommand{\be}{\begin{equation}}
\newcommand{\ee}{\end{equation}}

\newcommand{\tr}{\mbox{\rm tr}\,}

\def\la{\langle}
\def\ra{\rangle}

\newcommand{\riemgz}{g_0}

\renewcommand{\hbar}{{\overline \riemgz}}

\newcommand{\Mstar}{{M^*}}

\newcounter{mnotecount}[section]

\renewcommand{\themnotecount}{\thesection.\arabic{mnotecount}}

\newcommand{\mnote}[1]
{\protect{\stepcounter{mnotecount}}$^{\mbox{\footnotesize
$
\bullet$\themnotecount}}$ \marginpar{
\raggedright\tiny\em
$\!\!\!\!\!\!\,\bullet$\themnotecount: #1} }

\newcommand{\ptc}[1]{\mnote{{\bf ptc:}#1}}

\renewcommand{\H}{\mathcal H}

\newcommand{\bm}[1]{\mbox{\boldmath $#1$}}
\newcommand{\ov}[1]{\bar{#1}}

\newcommand{\xib}{\bm{\xi}}
\renewcommand{\a}{\alpha}
\renewcommand{\b}{\beta}
\renewcommand{\c}{\gamma}
\renewcommand{\d}{\delta}
\newcommand{\hOmega}{\widehat{\Omega}}

\newcommand{\Rsig}{(R^{h})}
\newcommand{\Csig}{(C^{h})}
\newcommand{\grad}{\mathrm{grad}}

\renewcommand\r{\widehat{r}}

\newcommand\B{\widehat{\Xi}}
\newcommand{\C}{\widehat{\Upsilon}^{(a)}}

\renewcommand{\ptcheck}[1]{}

\begin{document}
\title{On staticity of bifurcate Killing horizons\thanks{Vienna preprint  UWThPh 2023-12}
}
\author{
Piotr T. Chru\'sciel\thanks{Email \protect\url{piotr.chrusciel@univie.ac.at}, URL \protect\url{http://homepage.univie.ac.at/piotr.chrusciel/}}
\\
Faculty of Physics,
University of Vienna
\\
\phantom{x}
\\
Marc  Mars\thanks{Email \protect\url{marc@usal.es}}
\\
Institute of Fundamental Physics and Mathematics \\
University of Salamanca}
\date{}

\maketitle


\begin{abstract}
  We show that bifurcate Killing horizons with closed torsion form, in spacetimes of arbitrary dimension satisfying a Ricci-structure condition, arise from static Killing vectors. The result applies in particular to $\Lambda$-vacuum spacetimes.
\end{abstract}

\tableofcontents

\section{Introduction}

Stationary spacetimes are of particular physical importance as they describe equilibrium configurations of the gravitational field.  The static subcase, where the spacetime is not only time independent but also invariant under time reversal, plays a particularly important role. It is therefore of interest to identify conditions that guarantee that a stationary spacetime is   static. Recall that a stationary spacetime with a Killing vector field $\xi$ is \emph{static} if the twist form
\begin{equation}\label{22III23.11}
  \omega := \frac 12  \xib   \wedge d\xib
  \,,
\end{equation}
where $\xib$ is the $1$-form $g(\xi,\cdot)$,
vanishes; such Killing vectors will also be called \emph{static}. In situations where the spacetime is globally hyperbolic and the Killing vector $\xi$ is timelike everywhere it is immediate to show that the spacetime is static if and only if the twist form vanishes on a Cauchy surface. This is a consequence of the fact that $\omega$ is always Lie constant along the Killing  field. However, in black hole configurations the Killing field is not timelike everywhere and finding necessary and sufficient conditions for staticity is more delicate.

One of the central notions in the study of stationary black hole spacetimes is that of Killing horizons~\cite{ChBlackHoles}. Large families of examples have been constructed in \cite{KroenckePetersen,GerochHartle,RaczHorizons1,RaczHorizons2}; see also~\cite{ChPaetzKIDs}. This raises the question of  staticity in terms of the properties of Killing horizons. In fact, the  vanishing -- or not -- of the twist form plays an important role in the classification of stationary vacuum black holes, cf.\ e.g.~\cite{ChCo,CCH} and references therein.

 The aim of this work is to give a characterisation of static spacetimes containing a bifurcate Killing horizon in terms of the data on the horizon.
 We note  that $\omega$ always vanishes on a Killing horizon $\mcH$, so a criterion on a Killing horizon must involve at least one transverse derivatives on $\mcH$.
Further, both $\omega$ and its  first derivatives  vanish at the bifurcation surface of a bifurcate Killing horizon, so a criterion there must involve at least two transverse derivatives of $\omega$. The relevant fields are a priori not clear.

 We aim at finding such a characterization independently both of the matter fields involved and of the field equations of the theory, i.e. a purely geometry characterization valid in full generality. Given that no field equations are imposed, it is clear that such a characterization will necessarily involve two types of conditions, namely a bulk condition at the level of curvature that holds everywhere and a boundary condition that holds on the horizon. In this paper we deal with bifurcate Killing horizons, so it is plausible to expect that the boundary conditions can be formulated directly on the bifurcation surface. We find necessary and sufficient conditions for staticity both for the bulk and for the boundary conditions. The bulk condition involves a one-form constructed from the Ricci tensor and the Killing field. Specifically, it demands that, on all points where the Killing vector is not null, the one-form, denoted by
$g(\xi,\xi) r$, and
defined as the tangential-normal components of the Ricci tensor along the vector $\xi$,  is closed. The boundary condition at the bifurcation surface is that the \emph{torsion one-form} is closed.
 
It turns out that the analysis in the region to the causal future of the bifurcation surface of the bifurcate Killing horizon, and in its complement, is essentially different. This is related to the different causal character of the Killing orbits in these regions. The timelike-orbits region is analysed in Section~\ref{ss18VIII23.1}, while the space-like-orbits one in Section~\ref{ss24III23.1}.
The precise statements of our staticity results appear  in Theorems~\ref{t4XII21.1} and \ref{T23XII21.1} below. In addition to being independent of any field equations, our results hold  in any spacetime dimension and for any topology of the bifurcation surface. 

An interesting by-product of our just-mentioned staticity theorems
 is that whenever the torsion one-form of the bifurcation surface is closed, the fact that the bulk one-form $g(\xi,\xi) r$ is closed implies  that $\xi$ is an eigenvector of the Ricci tensor (equivalently, that  the one-form
  $g(\xi,\xi) r$ being closed  implies that it vanishes everywhere, provided the torsion form of the bifurcations surface is closed). 

Of course, of particular importance is the case where the spacetime is vacuum. In this case, the characteristic initial data for the spacetime on the bifurcate Killing horizon consists of  the metric induced on
the bifurcation surface $S$ together with the torsion $1$-form of $S$.
 A special case of our results  is the following theorem, where a non-vanishing cosmological constant is allowed,
with $\mcH^-$ denoting the part of $\mcH$ lying to the past of the bifurcation surface:

\begin{theorem}
  \label{t22III23.1}
  Consider a
  $(n+1)$-dimensional vacuum spacetime $(M,g)$, $n\ge 2$,
   containing a bifurcate Killing horizon $\mcH$. Let  $\mcD^-(\mcH^-)$ be the past domain of dependence of $\mcH^-$ in $M$ 
   and let $ \mathring M$
denote that connected component of the set 
  \begin{equation}\label{22III23.31}
    \mcD^-(\mcH^-)\cup \{g(\xi,\xi)<0\}
  \end{equation}
which contains $\mcD^-(\mcH^-)$.
  Then $g$  is static on $\mathring M$ if and only if the torsion $1$-form of the bifurcation surface is closed.
\end{theorem}

 \begin{remark}
 {\rm
As already said, we show in fact that  a   necessary and sufficient condition for  staticity  is closedness of the torsion together with
\begin{equation}
 d\big( g(\xi,\xi) \hat{r} \big) =0
 \,,
  \label{29III23.1}
\end{equation}
where  $\hat r$ is the quotient-space equivalent of the covector field $r_\alpha dx^\alpha $ appearing in the decomposition \eqref{22III23.22} of the Ricci tensor.
 This condition will of course hold in vacuum, but it will also hold  for, e.g., a perfect fluid supported in the region exterior to a  black hole and comoving with the Killing flow.
 \hfill $\Box$
 }
 \end{remark}

 \begin{remark}
 {\rm
 Our analysis below likewise shows that a vacuum spacetime metric will be static if and only if \eqref{29III23.1} holds and the first derivatives of $\omega$ vanish on one of the branches of a bifurcate horizon $\mcH$. This statement, however, does not seem to have an obvious formulation in terms of characteristic Cauchy data on $\mcH$.
 \hfill $\Box$
 }
 \end{remark}

 At the heart of the proof of Theorem~\ref{t22III23.1} lies the derivation of an equation for a relevant field, to which the unique continuation results of Mazzeo~\cite{Mazzeo-continuation} and the uniqueness results of Fuchsian symmetric hyperbolic equations apply.
Another ingredient is a  calculation tying the twist of the Killing vector to the torsion $1$-form of the horizon.
In order to finish the proof we need to treat separately the region $\mcD(S)$, where the Killing vector is spacelike, and
its complement in $\mathring M$.
Our claim on $\mcD(\mcH)$ is  obtained from uniqueness of solutions of Fuchsian hyperbolic equations;  this is the vacuum case of Theorem~\ref{T23XII21.1} below.
 The result in the complement of $\mcD(\mcH)$ in $\mathring M$ is the vacuum case of Theorem~\ref{t4XII21.1} below; it is based on the observation, which might have some interest of its own (cf.~Proposition~\ref{P22III23.1} below), that the bifurcation surface of a Killing horizon can be viewed as the conformal boundary at infinity of a natural rescaling of the quotient space-metric.

We note that in real-analytic spacetimes the result can be established by Taylor expansion at the horizon; this requires a similar computational effort. We further note that while a vacuum  spacetime metric will be real analytic near a Killing horizon in many situations of interest~\cite{BeigChMars,ChAPP},
 Theorems~\ref{t4XII21.1} and \ref{T23XII21.1} neither assume vacuum nor analyticity.

 \section{The twist form and the quotient space}
 \label{s18XII21.2}

 The aim of this section is to derive the key equation \eqref{20XII21.11-1} below,  satisfied by a field closely related to the twist of the Killing vector.

Let thus $\xi$ be a Killing vector field (KV) in an 
  $(n+1)$-dimensional spacetime
$(  \redM ,g)$, $n\ge 2$. 
 We use $\nabla$ for the Levi-Civita derivative of $g$. Let $\xib$ be the metrically associated $1$-form,
and define the $2$-form $F:= \frac{1}{2} d \xib$; recall that  the
{\it twist} $3$-form $\omega$ has been defined as $ \frac{1}{2} \xib \wedge d \xib$. In abstract index notation, 
\begin{align}
  F_{\a\b}= \nabla_{\a} \xi_{\b}, \qquad
  \omega_{\a\b\c} = \xi_{\a} \nabla_{\b} \xi_{\c}
  +\xi_{\b} \nabla_{\c} \xi_{\a}
  +\xi_{\c} \nabla_{\a} \xi_{\b}.
  \label{defF_omega}
\end{align}
Recall that $\lambda := - g(\xi,\xi)$ and define the $2$-form $\Omega:=i_{\xi} \omega$.
Note that, by construction, $\Omega$ is Lie-constant along $\xi$ and
orthogonal to $\xi$. It is immediate that
\begin{align}
  \Omega = -\lambda F - \frac{1}{2} \xib \wedge d \lambda.
  \label{F}
\end{align}

Define
$$\Mstar  := \{ p \in  \redM  ; \lambda (p) \neq 0\}
 \,,
$$
i.e. the open set where
$\xi$ is non-null, and assume it to be non-empty. From the definitions of $\omega$, $F$, and \eqref{F} we have
\begin{align}
  \omega = - \frac{1}{\lambda} \xi \wedge  \Omega \qquad  \mbox{on } \quad \Mstar
   \,.
    \label{24III23.31}
\end{align}
Note that the left-hand side is smooth on $ \redM $, which implies that   the right-hand side extends smoothly across the zero-level set of $\lambda$. In particular we have:
\begin{lemma}
  \label{SufficiencyOmega}
  If $\Mstar $ is dense in $ \redM $ then the vanishing of $\Omega$ is equivalent to the vanishing of $\omega$.
  \end{lemma}

Using  $F = \frac{1}{2} d \xib$, the expression \eqref{F} can  be written as
\begin{align*}
  \Omega = - \frac{\lambda^2}{2}
  d \left ( \lambda^{-1} \xib \right ), \qquad \mbox{on } \quad \Mstar
\end{align*}
from which it immediately follows that
\begin{align}
  d \left ( \lambda^{-2}  \Omega \right ) =0, \qquad \mbox{on } \quad \Mstar .
  \label{dOmega}
\end{align}
  This is equation (B.10) in Appendix B of \cite{ACD2}. The Killing vector $\xi$ satisfies the standard identity ($R^{\a}{}_{\b\c\d}$ is the Riemann tensor of $g$)
  \begin{align*}
    \nabla_{\a} \nabla_{\b} \xi_{\c} = \xi_{\sigma} R^{\sigma}{}_{\a\b\c}.
  \end{align*}
  Directly from \eqref{defF_omega} (brackets denote antisymmetrisation)
  \begin{align*}
    \nabla_{\mu} \omega_{\a\b\c} =
    F_{\mu\a} F_{\b\c} + F_{\mu\b} F_{\c\a} +  F_{\mu\c} F_{\a\b}
    + 3 \xi_{\sigma} R^{\sigma}{}_{\mu[\a\b} \xi_{\c]}.
  \end{align*}
  Contracting $\mu$ and $\a$ yields $(R_{\alpha\beta}$ is the Ricci tensor of $g$)
  \begin{align}
    \nabla^{\mu} \omega_{\mu\b\c}  = \xi_{\sigma} \left (
    \xi_{\b} R^{\sigma}{}_{\c} -
    \xi_{\c} R^{\sigma}{}_{\b}  \right ).
    \label{divomega}
  \end{align}
  This corresponds to (3.20) in Appendix B of \cite{ACD2} and recovers the well-known fact that
  $\omega$ is co-closed when $\xi$ is an eigenvector of the Ricci tensor.

  We wish to compute the codifferential of $\Omega$ on the orbit space of $\xi$. We assume that the space $\Sigma := \Mstar  / \sim$  is a smooth manifold,
  where we write that $p \sim q$ if and only if the points $p, q \in \Mstar $ belong to the same integral curve of $\xi$.
It is well-known that
  covariant tensors on $\Mstar $ which are Lie-constant along $\xi$
  and totally orthogonal to
  $\xi$ descend to the quotient. We use a hat to denote the corresponding object on $\Sigma$. We only break this rule for scalars, such as
   \begin{equation}\label{24III23.1}
   \lambda := - g(\xi,\xi)
\end{equation}
 and
  for the tensor
   \begin{equation}\label{24III23.2}
    h_{\a\b} := g_{\a\b} + \lambda^{-1} \xi_{\a} \xi_{\b}
     \,,
   \end{equation}
   which descends to a metric on $\Sigma$. In these last two cases we use
  the same symbol for the spacetime and for the quotient objects.

  The operation
  \begin{align*}
    D_{\a} T_{\b_1 \cdots \b_p} := h^{\c}{}_{\a}
    h^{\d_1}{}_{\b_1} \cdots h^{\d_p}{}_{\b_p} \nabla_{\c} T_{\d_1 \cdots \d_p}
  \end{align*}
  on any $\xi$-Lie-constant and $\xi$-orthogonal covariant tensor $T$, gives a tensor with the same properties. It therefore induces an operation on $\Sigma$. It is well-known that this operation is the Levi-Civita covariant derivative
  with respect to $h$.
  We also use $D$ for this derivative on $\Sigma$.

  We already noted that $\Omega$ is $\xi$-Lie-constant and $\xi$-orthogonal. Let $\hOmega$ be the (uniquely defined) corresponding two-form on $\Sigma$:
 if we denote by $\pi$ the projection from $M^*$ to $M^*/\sim$, then
   \begin{equation}
    \label{20III23.1}
     \Omega = \pi^* \hat \Omega\,.
   \end{equation}

  Exterior differentiation commutes with the operation of descending to the quotient for forms that descend (this is clearly seen in local coordinates adapted to the isometries),
   so \eqref{dOmega} implies
  \begin{align*}
d \left ( \lambda^{-2}  \hOmega \right ) =0.
  \end{align*}
  With the aim of computing the codifferential of $\widehat{\Omega}$ we consider an arbitrary $\xi$-Lie-constant $\xi$-orthogonal $2$-covariant tensor $P_{\a\b}$ and compute
  \begin{align*}
    D_{\mu} P^{\mu}{}_{\b} & = g^{\mu\a}
    D_{\mu} P_{\a\b} = h^{\mu\a} h^{\c}{}_{\b} \nabla_{\mu} P_{\a\c}
    = h^{\c}{}_{\b} \left ( \nabla^{\mu} P_{\mu\c}
    + \frac{1}{\lambda} \xi^{\mu} \xi^{\a} \nabla_{\mu} P_{\a\c} \right ) \\
                           & = h^{\c}{}_{\b} \nabla^{\mu} P_{\mu\c}
                             - \frac{1}{2 \lambda} P_{\mu\b} \nabla^{\mu} \lambda.
  \end{align*}
  Let us define the covector
  \begin{align}
    r_{\b} := \lambda^{-2} \xi^{\a} h^{\c}{}_{\b} \nabla_{\mu} \omega^{\mu}{}_{\a\c}.
 \label{20III23.21}
  \end{align}
  Then $r$ is orthogonal to $\xi$ and invariant under the flow of $\xi$, hence there exists a unique covector field $\hat r$ on the quotient $M^*/\sim$ such that
  \begin{equation}\label{20III23.2}
    r = \pi^* \hat r
    \,.
  \end{equation}

  On $\Mstar $ we can decompose the Ricci tensor in components orthogonal to $\xi$. Using  \eqref{divomega} it follows that the normal-tangential component
  of the Ricci tensor is precisely $r_{\a}$. In other words, the decomposition takes the form
  \begin{align}
    R_{\a\b} = r \xi_{\a} \xi_{\b} + r_{\a} \xi_{\b} + r_{\b} \xi_{\a} + r_{\a\b}
   \quad  \mbox{on} \ \Mstar
   \,,
    \label{22III23.22}
  \end{align}
  where $r_{\a\b}$ is symmetric and orthogonal to $\xi$.
  It follows that $r_{\a}$ vanishes if and only if $\xi$ is an eigenvector of Ricci on $\Mstar $.
  Thus
  \begin{align*}
    D_{\mu} \Omega^{\mu}{}_{\b} &=
h^{\c}{}_{\b}    \nabla_{\mu} \Omega^{\mu}{}_{\c}
    - \frac{1}{2 \lambda} \Omega_{\mu\b} \nabla^{\mu} \lambda
    = \nabla_{\mu} \left ( \xi^{\a} \omega_{\a}{}^{\mu}{}_{\c} \right )
    h^{\c}{}_{\b}
    - \frac{1}{2 \lambda} \Omega_{\mu\b} \nabla^{\mu} \lambda \\
    & = (\nabla_{\mu} \xi^{\a}) \omega_{\a}{}^{\mu}{}_{\c}
      h^{\c}{}_{\b}
     + \xi^{\a} h^{\c}{}_{\b}  \nabla_{\mu} \omega_{\a}{}^{\mu}{}_{\c}
      - \frac{1}{2 \lambda} \Omega_{\mu\b} \nabla^{\mu} \lambda \\
                                & = - h^{\c}{}_{\b} \omega_{\a\mu\c} \nabla^{\a} \xi^{\mu}
                                  - \lambda^2 r_{\b}
                                  - \frac{1}{2 \lambda} \Omega_{\mu\b} \nabla^{\mu} \lambda.
                                   \end{align*}
Now, on $\Mstar $ the expression \eqref{F} allows us to write $\nabla_{\a}\xi_{\mu}$ in terms of $\Omega_{\a\mu}$ as
\begin{align}
\nabla_{\a} \xi_{\mu} = - \frac{1}{\lambda}
\left ( \Omega_{\a\mu} + \frac{1}{2} \xi_{\a} \nabla_{\mu}\lambda
- \frac{1}{2} \xi_{\mu} \nabla_{\a}\lambda \right )
 \,.
\label{F2}
\end{align}
Inserting this into the before-last equation yields
\begin{align}
    D_{\mu} \Omega^{\mu}{}_{\b} &= - \lambda^2 r_{\b}
+ \frac{1}{2 \lambda} \Omega_{\mu\b} \nabla^{\mu} \lambda
+ \frac{1}{\lambda} \omega_{\a\mu\c} \Omega^{\a\mu} h^{\c}{}_{\b}.
\label{intermediate}
\end{align}
The last term is zero because
$\Omega^{\a\mu} h^{\c}{}_{\b}$  is orthogonal to $\xi$ in all contravariant indices, while $\omega = \xib \wedge F$ (see \eqref{defF_omega}).

We conclude that the differential
and codifferential of $\hOmega$ on $\Sigma$ satisfy
\begin{align}
 d \left ( \frac{1}{\lambda^{2}} \hOmega \right ) =0, \qquad
 \delta \left (\frac{1}{| \lambda|^{\frac{1}{2}}}  \hOmega \right )
 =  |\lambda|^{\frac{3}{2}} \r
 \,,
  \label{28XI21.1p}
\end{align}
where we use the sign convention for the codifferential $\delta$ as in
\cite{deRahm}:
\begin{equation}\label{20III23.5}
 (\delta \sigma)_{b_2 \cdots b_p} = - D^{b_1} \sigma_{b_1 b_2 \cdots b_p}
\,.
\end{equation}

Our next task is to derive an identity
for $\Delta_h \hOmega$ using the Weizenb\"ock formula. Define the two-form
\begin{equation}\label{22III23.3}
 \B := \lambda^{-2} \hOmega
 \,.
\end{equation}
It will be shown in Section~\ref{s28XII21.1}  that $\B$ extends continuously to that part of the boundary of the quotient space which corresponds to the Killing horizon, and is related to the torsion $1$-form of the bifurcation surface.
We will also show  that
$\B$ vanishes if and only the twist of the Killing vector vanishes.

In order to obtain the promised identity we apply the trivial identity $\delta ( f \alpha) = f \delta \alpha - i_{ \grad \ f} \alpha$, valid form any scalar $f$ and $p$-form $\alpha$, and obtain
\begin{align}
  \delta  \B =
  \r + \frac{3}{2 \lambda} i_{\grad \lambda} \B.
  \label{deltaB}
\end{align}
Since  $d \B=0 $, we can easily compute the Hodge-de Rahm operator
$\Delta_{\mbox{\tiny dR}}
:= (d \circ \delta + \delta \circ d)$ of $\B$
\begin{align*}
  \Delta_{\mbox{\tiny dR}} \B = (d \circ\delta) \, \B =
  d \left ( \r
  + \frac{3}{2} i_{\grad(\ln |\lambda|)} \B \right ) = d  \r  + \frac{3}{2} \pounds_{\grad(\ln |\lambda|)} \B,
\end{align*}
where in the last equality we used the Cartan identity $\pounds_X = i_X \circ d + d \circ i_X$. We now apply the Weizenb\"ock identity \cite{Weizenbock,deRahm},
which for $2$-forms reads
\begin{align}
  (\Delta_h \B)_{ij}= -  (\Delta_{\mbox{\tiny dR}} \B)_{ij}
  - 2 \Rsig^{k}{}_{i}{}^{\ell}{}_j
  \B_{kl} + \Rsig^{k}{}_i \B_{kj}
  - \Rsig^k{}_j \B_{ki} \label{Weiz}
\end{align}
where $\Delta_h$ is the rough Laplacian $\Delta_h := D^i D_i$
and the Riemann and Ricci tensors of $(\Sigma,h)$ are denoted respectively by
$\Rsig^{i}{}_{jkl}$ and $\Rsig_{ {j\ell }}$. Thus, the field $\B$ satisfies the identity
\begin{align}
\Delta_h \B_{ij}
= & - 2 \Rsig^{k}{}_{i}{}^{\ell}{}_j
  \B_{kl} + \Rsig^{k}{}_i \B_{kj}
  - \Rsig^k{}_j \B_{ki}
  - D_i   \r_j
  + D_j   \r_i  \nonumber \\
  &    - \frac{3}{2} \pounds_{\grad(\ln |\lambda|)} \B_{ij}.
    \label{IdentityB}
\end{align}

Recall that  $-g(\xi,\xi) \equiv \lambda = \epsilon u^2$, $\epsilon = \pm 1$, with  $u>0$.
It will be useful to write down an equivalent identity involving
the two-form
\begin{equation*}
 \C := u^a \, \B = u^{a-4} \hOmega
 \,,
 \quad
  a \in \mathbb{R}
  \,,
\end{equation*}
where the subscript $a$ on $\C$ is \emph{not} a vector index, but indicates the power $a$ used in the rescaling.
We are mainly interested in the case
\begin{equation}\label{22III23.2}
  a = a_n:=\frac{n-1}{2}
   \,,
\end{equation}
but we leave the constant $a$ arbitrary unless explicitly specified otherwise.

A straightforward computation gives
\begin{align}
  d \C & = \frac{a}{u} du \wedge \C, \label{dC}\\
  \delta \C & = u^a \r + (3-a) i_{\grad \ln u} \C, \label{deltaC} \\
    \Delta_h \C_{ij}  = &
  - 2 \Rsig^{k}{}_{i}{}^{\ell}{}_j
  \C_{kl} + \left ( \Rsig^{k}{}_i - \frac{3 D_i D^k u}{u} + \frac{ 3 D^k u D_i u}{u^2}
  \right )   \C_{kj} \nonumber \\
  & -
  \left ( \Rsig^k{}_j - \frac{3 D_j D^k u}{u} + \frac{3 D^k u D_j u}{u^2} \right )
  \C_{ki}
  - 2 u^a D_{[i} \left ( \r_{j]} \right ) \nonumber \\
 &
  + (2 a -3) \frac{D^k u}{u} D_k \C_{ij}
  + a \left ( \frac{\Delta_h u}{u} + (2-a) \frac{ |d u|^2_h}{u^2}
  \right ) \C_{ij}. \label{identity}
\end{align}
It will also be of interest to find an equation in terms of a conformal metric. Define
\begin{equation}\label{24III23.81}
 \ov{h}_{ij} := u^{-2} h_{ij}
 \,.
\end{equation}
Applying Lemma \ref{transConf} in
Appendix \ref{ApA} to $\Psi = \C$ and using \eqref{dC}-\eqref{deltaC} we find
\begin{align*}
  \Delta_{\ov{h}} \C_{ij}
     = & u^2 \Delta_h \C_{ij} + 2 \left ( u \Delta_h u + (2 +a -n) |d u|^2_h
    \right ) \C_{ij} \\
    & + (4-n) u D^k u D_k \C_{ij}
- (n+2) D^k u D_i u \, \C_{kj}
    + (n+2) D^k u D_j u \, \C_{ki}
    \\
    & -2 u^{1+a} \left ( D_i u \, \r_j - D_j u \, \r_i \right ).
\end{align*}
Combining with \eqref{identity} we conclude that $\C$ satisfies
\begin{align}
 L_a (\C) = - u^a d (u^2 \r)
 \,, \label{mainEq}
\end{align}
where $L_a$ is the operator
\begin{align}
L_a(\Psi)_{ij} := &
\Delta_{\ov{h}} \Psi_{ij}
+ (n-1-2a) u D^k u \, D_k \Psi_{ij}
 \nonumber
\\
 &
+ 2 u^2 \Rsig^k{}_i{}^{\ell}{}_j \Psi_{kl}
+ B^k{}_i \Psi_{kj}
- B^k{}_j \Psi_{ki}
 \,,
 \label{20XII21.11-1}
 \\
B^{k}{}_i  := &
- u^2 (\Rsig)^k{}_i + 3 u D^k D_i u + (n-1) D^k u D_i u
 \nonumber
\\
& - \left ( \frac{2+a}{2} u \Delta_h u
+ \big ( (2 +a -n ) + \frac{a(2-a)}{2} \big)
|d u|^2_h \right ) \delta^k{}_i
 \,.
 \label{20XII21.11}
\end{align}

\begin{lemma}
 \label{l18XII21.1}
The operator $L_a$ with $a = \frac{n-1}{2}$ is formally self-adjoint with respect to the $L^2(d\mu_{\ov{h}})$-scalar product, where $d\mu_{\ov{h}}$ is the Riemannian measure defined by $\ov{h}$.
\end{lemma}

{\noindent \sc Proof:}
For this value of $a$  there are no first order terms in $L_a$. The symmetry of the zero-order terms  is the contents of
Corollary \ref{self} in Appendix \ref{ApA}.
$\hfill \Box$
\bigskip

\section{R\'acz-Wald coordinates}
 \label{s28XII21.1}

We wish to tie the quantities introduced above to objects defined on the bifurcation surface of a bifurcate Killing horizon. For this a convenient
set of coordinates is the one due to  R\'acz and Wald~\cite{RaczWald1,RaczWald2}. Such coordinates can be introduced in either of the following circumstances:

\begin{enumerate}
  \item[H1.] The spacetime contains a bifurcate Killing horizon.
  \item[H2.] The spacetime has a   null boundary $\mcN$ with a causal Killing vector $\xi$ which is tangent to the generators of $\mcN$, with non-zero surface gravity. The orbits of $\xi$ are complete and there exists a hypersurface $\hyp$ to which $\xi$ is transverse and such that every orbits of $\xi$ intersects $\hyp$ precisely once.
\end{enumerate}

In the latter case, which is typical for well-behaved domains of outer communications, one constructs~\cite{RaczWald1} an auxiliary spacetime which contains a bifurcate Killing horizon. So, in this sense, it suffices to consider   spacetimes $( \redM ,g)$ which contain a bifurcate Killing horizon, i.e. two smooth null connected hypersurfaces
$\H^+$ and $\H^-$ intersecting on a spacelike codimension-two surface
$S = \H^+ \cap \H^-$ and such that $(  \redM ,g)$ admits a non-trivial
Killing vector $\xi$ which vanishes on $S$. It follows automatically that
$\H^{\pm} \setminus S$ are Killing horizons of $\xi$ with two connected components each.

A spacetime neighbourhood $\mcO$ 
 of $S$ can be coordinatised with an atlas of
\emph{R\'acz-Wald coordinates}
$\{ U,V, x_{\mathfrak{i}}^a \}$, where $\{ x^a_{\mathfrak{i}}, O_{\mathfrak{i}} \}$ is an atlas of $S$. In such coordinates, the metric $g$ takes the form
\begin{align}
  g = G dU \left ( dV + V q_a dx^a \right ) + \gamma_{ab} dx^a dx^b
  \,,
  \label{RWmetric}
\end{align}
where all metric coefficients $G,q_a, \gamma_{ab}$ depend only upon $\{UV, x^a\}$,
and where $G$ is a positive function. The bifurcation surface $S$ is the set
  $\{U = V =0 \}$.
  We choose the time orientation so that $\partial_U$ is future pointing and $\partial_ V$ is past pointing. The construction of~\cite{RaczWald1} guarantees that there exists a neighborhood of the bifurcate Killing horizon on which $G$ is strictly positive.

  Recall that the torsion $1$-form of a spacelike surface of  codimension equal to two is defined as
  \begin{align*}
    \zeta(X) =  g ( k, \nabla_X \ell)
  \end{align*}
  where $X$ is a vector field in $S$ and $\{k,\ell\}$ are a pair of null normals to $S$ satisfying the normalization condition $g(k,\ell) = -1$. The null normals  $k$ and $\ell$ can be boosted or interchanged. The boost transforms the torsion by adding  the differential of a function,
   while the interchange $k \leftrightarrow \ell$ changes the sign of $\zeta$. Note that the property of $\zeta$ being closed if not affected by these changes.

   In what follows we shall need the explicit form of $\zeta$. We take the normal basis
  \begin{align*}
    k =- \sqrt{\frac{2}{G}} \partial_V\,, \qquad   \ell = \sqrt{\frac{2}{G}}
    \partial_U
    \,,
  \end{align*}
  and after a simple calculation based on the metric \eqref{RWmetric} we find
  \begin{align}
    \zeta = \frac{1}{2} q_a |_{\{UV=0\}} dx^a
    \,.
     \label{RWtorsion}
\end{align}

In R\'acz-Wald coordinates
the Killing vector $\xi$ reads
\begin{align}
  \xi = U \partial_U  - V \partial_V. \label{RWKilling}
\end{align}
and therefore
\begin{align}
  \xib:= g(\xi, \cdot) = \frac{1}{2} G \left ( U dV - V dU + UV  q_{a} dx^a \right )
  \,.\label{RWxib}
\end{align}

We shall also need the explicit form of the twist $3$-form of $\xi$, recall that
$$\omega
 := \frac{1}{2} \xib \wedge d \xib
  \,.
$$
For this it is simplest to work away from the bifurcate Killing horizon and write $\xib$ as
 \ptcheck{23III23}
\begin{align*}
\xib = \frac{1}{2} G UV \left ( d \ln \left |\frac{V}{U} \right | - q_a dx^a \right )
 = :
\frac{1}{2} G UV \tilde{\xib}.
\end{align*}
Since $q_a=q_a(UV,x^b)$, a direct calculation gives, where a prime denotes a derivative with respect to $UV$:
\begin{align*}
\tilde{\xib} \wedge d \tilde{\xib} = &
 2 q'_c dU \wedge dV \wedge dx^c +
 q'_a q_c \left ( U dV + V dU \right )  \wedge dx^a \wedge dx^c
\\
 &
 -
  \left ( \partial_a q_b - \partial_b q_a \right ) \left (
\frac{dV}{V} - \frac{dU}{U} \right ) \wedge dx^a \wedge dx^b
 + q_c (\partial_a q_b - \partial_b q_a ) dx^c \wedge dx^a \wedge dx^b
 \end{align*}
and we conclude that
 \ptcheck{24III}
\begin{align*}
  \omega = \,
   &
   \frac{1}{8} G^2 U^2 V^2 \tilde{\xib} \wedge d \tilde{\xib}
   \nonumber
\\
   = \,
    &
    - \frac{1}{8} G^2 \left ( \partial_a q_b - \partial_b q_a \right ) \left (
U^2 V dV  - V^2 U dU \right ) \wedge dx^a \wedge dx^b
+ U^2 V^2 \widetilde{\omega}
 \,,
\end{align*}
where $\widetilde{\omega}$ is a smooth $3$-form. Note that although the computation has been done away from $U=0$ or $V=0$, the result is valid everywhere. Observe also  that the twist always vanishes on the bifurcate Killing horizon. This is a consequence of the fact that $\xi$ is a null normal to the hypersurfaces $U=0$ and $V=0$.

The transverse derivative $\pounds_{\partial_V} \omega$ of $\omega$ at the null hypersurface $\{ V=0\}$ reads
 \ptcheck{24III}
\begin{align}
  \pounds_{\partial _V} \bm{\omega} |_{\{ V=0\}} =
  -
  \frac{1}{8} G^2  \left ( \partial_a q_b - \partial_b q_a \right ) |_{UV=0}
  U^2 dV\wedge dx^a \wedge dx^b.
  \label{24III23.85}
\end{align}
This $3$-form vanishes if and only if $q_a |_{\{UV=0\}}$ is closed, i.e. if and only if the torsion $1$-form $\zeta$ of the bifurcation is closed. A similar statement  holds for the null hypersurface $\{ U=0\}$.

Recall that
$$\lambda  = - g(\xi,\xi)
$$
denotes the Lorentzian norm-squared of the Killing vector $\xi$, and that we write the \emph{quotient-space metric} as
$$
 h_{\alpha\beta} \equiv g_{\alpha\beta} + \lambda^{-1} \xi_{\alpha} \xi_{\beta}
  \,.
$$
Since $\lambda = - \xib(\xi) =  UV G$ a straightforward computation gives
\begin{align*}
  h = \frac{G}{4 UV} \left ( (U dV + VdU)^2 + U^2 V^2 (q_{a}  dx^a)^2
  + 2 U V (U dV + V dU) q_a dx^a \right ) + \gamma_{ab} dx^a dx^b.
\end{align*}
We introduce the scalar
$$
 s := UV
$$
and note that $\{s, x^a\}$ descend to local coordinates on the quotient space (both are Lie-constant along $\xi$). The tensor $h$ takes the form
\begin{align*}
  h = \frac{G}{4 s} \left ( ds + s q_a dx^a \right )^2
  + \gamma_{ab} dx^ a dx^b, \qquad  \mbox{and}\ \lambda =  s G
\end{align*}
where $G,q_a, \gamma_{ab}$ are functions of $\{ s, x^a\}$. Let us perform the coordinate change
\begin{equation*}
 s = \epsilon z^2
 \,,
 \quad
  \epsilon = \pm 1\,, \ z >0
  \,,
\end{equation*}
and note that
$\epsilon = 1$ corresponds to the domain $UV >0$ where the Killing is timelike, while $\epsilon = -1$ corresponds to the domain $UV<0$, where it is spacelike.

In the coordinates $\{z,x^a\}$, the quotient metric is
\begin{align*}
  h = \epsilon G \big( dz + \frac{1}{2} z q_a  dx^a\big)^2
  + \gamma_{ab} dx^a dx^b, \qquad \lambda = \epsilon z^2 G
  \end{align*}
where $G,q_a, \gamma_{ab}$ are functions of $\{ \epsilon z^2, x^a \}$.
This defines a metric smooth-up-to the boundary $z=0$, because $G$ is positive and $\gamma_{ab}$  positive
  definite in the domain $\mcO$. 
   Clearly $h$ is Riemannian when $\epsilon>0$, i.e. in the region where the Killing vector field $\xi$ is timelike, and Lorentzian when $\epsilon<0$, where $\xi$ is spacelike.

Let $u >0 $ be defined by
\begin{equation}\label{24III23.3}
 \lambda = \epsilon u^2\,,
 \quad
  \epsilon = \pm 1
  \,.
\end{equation}
The function $u$ in the $\{z,x^a\}$ coordinates reads
  $u = z \sqrt{G}$, so it is smooth-up-to-and-including the boundary, where it has a zero of first order. Thus $u$ is a defining function
  for $\partial \Sigma$.
The metric
$$
\ov{h} = u^{-2} h
$$
appearing in \eqref{20XII21.11-1} provides therefore an example of smoothly conformally compactifiable metric \emph{\`a la Penrose}. In the region where the Killing vector $\xi$ is spacelike the bifurcation surface $S\approx \partial \Sigma$ becomes the conformal boundary at timelike infinity for the metric $\ov{h}$. This property will be exploited in what follows.

Perhaps unexpectedly, $|d u|^2_h$ is constant on $\partial\Sigma$.
To see this it is convenient to introduce $P := \sqrt{G}$ so that
  $u = z P$, and raise indices $a,b$ with $\gamma^{ab}$ defined to be the inverse of $\gamma_{ab}$. The inverse metric $h^{\sharp}$ is given by
  \begin{align*}
    h^{\sharp} = \left (
    \begin{array}{cc}
      \frac{\epsilon}{P^2} + \frac{1}{4} z^2 q^a q_a & - \frac{1}{2} z q^a \\
      - \frac{1}{2} z q^a & \gamma^{ab} \\
    \end{array}
    \right ).
  \end{align*}
  Thus
  \begin{align*}
    \frac{1}{z} \grad_h z = \frac{\epsilon}{z P^2} \partial_z + O(1),
    \qquad \frac{1}{u} \grad_h u =
    \frac{1}{z} \grad_h z +\frac{1}{P} \grad_h P =     \frac{\epsilon}{z P^2} \partial_z + O(1),
  \end{align*}
  where $O(1)$ denotes any function that extends smoothly to the boundary $\partial \Sigma$.
  Using the fact that $\partial_z$ acting on any metric coefficient $G,q_a, \gamma_{ab}$ gives a function of  the form $z F$, where $F$ is regular up to boundary (because the functions depend on $\{ \epsilon z^2, x\}$) it follows that
\begin{align*}
    \frac{1}{z} \left < \grad_h z , \grad_h P \right > = O(1)
     \,,
\end{align*}
where $\la \, , \, \ra$ denote scalar product with $h$. Consequently
    \begin{align*}
\frac{ \la \grad_h u, \grad_h u \ra}{u^2}
=  \frac{ \la \grad_h z, \grad_h z \ra}{z^2} + O(1)
      = \frac{ \epsilon}{z^2 P^2} + O(1) =
      \frac{\epsilon}{u^2} + O(1)
    \end{align*}
    and we find that
\begin{equation}\label{20XII21.1}
     |d u|^2_h |_{\partial \Sigma}=\epsilon
     \,.
\end{equation}
    This has the well known
    geometric interpretation  that

    \begin{proposition}
     \label{P22III23.1}
     In the region $\epsilon>0$ the metric $u^{-2}h$ is asymptotically locally hyperbolic, i.e., it is smoothly conformally compactifiable, with sectional curvatures tending to a negative constant when the conformal boundary at infinity $\partial \Sigma$ of $u^{-2}h$ is approached.
 \hfill $\Box$
 \end{proposition}

We finish this section with a formula for  the $2$-form $\B$ of \eqref{22III23.3} in terms of the R\'acz-Wald coordinates.
A computation of the two-form
$\Omega$ in the metric \eqref{RWmetric} gives
\begin{align}
  \Omega = - \frac{1}{4} G^2 (UV)^2 d q, \qquad q:= q_a dx^a
   \,.
    \label{22III23.5}
  \end{align}
  Hence  $\hOmega = - \frac{1}{4} u^4 dq$ and, in the coordinates $\{ z, x^a\}$,
\begin{equation}
  \B_{ij} = - \frac{1}{4}  \left ( \partial_i q_j - \partial_j q_i
  \right )
   \,,
  \qquad q_i := \delta^a{}_i q_a
  \,,
  \nonumber
\end{equation}
i.e.
\begin{equation}
  \B_{za}  = - \frac{1}{4} \partial_z q_a
  \,,
    \qquad\B_{ab} = - \frac{1}{4} \left ( \partial_a q_b - \partial_b q_a \right )
    \,.
\end{equation}
Since $q_a=q_a(z^2,x)$, all odd $z$-derivatives at $z=0$ vanish. Combining with the explicit form \eqref{RWtorsion} of the torsion $1$-form $\zeta$ of the bifurcation surface  we conclude that:

\begin{proposition}
 \label{p22III22.2}
  $\B_{ij} =0$ at $\partial \Sigma$ if and only  if the torsion $1$-form $\zeta$  of
$S$ is closed.
%
%
\hfill$\Box$
\end{proposition}

\section{Staticity}
 \label{28III23}

We are ready now to provide necessary and sufficient conditions for staticity of the spacetime metric. It turns out that the sides of the Killing horizons where the Killing vector is timelike and these where the Killing vector is spacelike require separate treatment.

\subsection{The timelike region}
 \label{ss18VIII23.1}

We start by considering the region where the Killing vector is timelike, which corresponds to $\epsilon =1$. However, all local calculations in this section apply as long as the Killing vector is not null, and will also be relevant for Section~\ref{ss24III23.1}.

    Using the notation
    of \cite{Lee:fredholm},
    it follows that $(\Sigma,\ov{h})$ has a compact conformal boundary at infinity and that $\Delta_{\ov{h}}$ is a uniformly degenerate   elliptic operator.
    We also use the notion of indicial map $I_{\sigma}$ as defined in
    \cite{Lee:fredholm}. Namely, given  a uniformly degenerate elliptic operator
    \begin{align*}
      L: \Gamma(E) \longrightarrow \Gamma(E)
    \end{align*}
    where $\Gamma(E)$ is the set of sections of a tensor bundle $E$ over  $\Sigma$,
    and using our function $u$ as the defining function for the boundary
     (which applies near horizons with non-zero surface gravity, but does not when the surface gravity vanishes),
     for $\sigma \in \mathbb{C}$
     the indicial map is the map
    \begin{align*}
      I_{\sigma} : \Gamma(E|_{\partial \Sigma} \otimes \mathbb{C}) & \longrightarrow
       \Gamma(E|_{\partial \Sigma} \otimes \mathbb{C} )\\
      \widehat{\Psi} & \longrightarrow I_{\sigma}(\widehat{\Psi}) := \lim_{u\to0} \frac{L ( u^\sigma \Psi) }{u^{\sigma}}
             \,,
    \end{align*}
    where $\Gamma(E|_{\partial \Sigma})$
    is the set of sections over
    $\partial \Sigma$ of the bundle $E$, $E|_{\partial \Sigma} \otimes \mathbb{C}$ is its complexification and
    $\Psi \in \Gamma(E \otimes \mathbb{C})$ is any smooth extension
    of $\widehat{\Psi}$. The indicial map is a well-defined (i.e. independent of the choice of extension $\Psi$),
    and is also independent of the choice of the defining function.
    The indicial exponents of $L$ are the
    values $\sigma$ for with there exists a point $p \in \partial \Sigma$
    and a non-zero $\widehat{\Psi}$
such that $I_{\sigma} (\widehat{\Psi}) |_{p} =0$.

We wish to prove a vanishing theorem for $\C$, for this we need to determine the indicial exponents for the operator~\eqref{mainEq}.

\begin{lemma}
\label{indicial}
  In the setup where $(  \redM ,g)$ admits a bifurcate Killing horizon, consider the
  quotient $\Sigma$ in the region where $\xi$ is timelike, i.e.
  $\epsilon =1$.
Define $L := L_a$ for $a= (n-1)/2$, cf.~\eqref{22III23.2}.
Then, the indicial
  exponents of $L$ are
\begin{align*}
 \sigma \in \left \{ a-4, a-3,
 a-1, a \right \}=
  \left\{ \frac{n-1}{2}-4, \frac{n-1}{2}-3,
\frac{n-1}{2}-1, \frac{n-1}{2}\right\}
 \,.
\end{align*}
\end{lemma}

{\noindent \sc Proof:}
Define (cf. Corollary \ref{self}, Appendix~\ref{ApA})
\begin{align*}
\ov{S}_{ij}{}^{kl} := 2 u^2 \Rsig^{[k}{}_i{}^{l]}{}_j
- \delta^{[k}{}_j B^{l]}{}_i - B^{[k}{}_j \delta^{l]}{}_{i}
\end{align*}
so that $L = \Delta_{\ov{h}} + S$.
We apply item (ii) in Lemma \ref{indicialmap}. Let
$$
V^i := D^i u |_{\partial \Sigma}
$$
be the restriction of the gradient of $u$ to  $\partial \Sigma$.
Given that
\begin{align*}
B^k{}_i |_{\partial \Sigma} =
  \big (
(n-1) V^k  V_i + \frac{n^2 -2 n -7}{8} |V|^2
\big)\delta^k_{i} \big|_{\partial_{\Sigma}}
\end{align*}
we have
\begin{align*}
  \ov{S}_{ij}{}^{kl} |_{\partial \Sigma} \widehat{\Psi}_{kl}
  = (n-1) \left ( V^k V_i \widehat{\Psi}_{kj} -
  V^k V_j \widehat{\Psi}_{ki} \right )
  + \frac{n^2 - 2 n -7}{4} \widehat{\Psi}_{ij}
  \end{align*}
and
the indicial map is
\begin{align*}
 I_{\sigma} (\widehat{\Psi})_{ij} = c_1 |V|^2 \widehat{\Psi}_{ij}
 + 3 \left ( V^k V_i\widehat{\Psi}_{kj} -
 V_k V_j \widehat{\Psi}_{ki} \right ), \quad
 c_1
  & :=  \left (\sigma - \frac{n -1}{2}  \right )
     \left ( \sigma - \frac{n-9}{2} \right ).
\end{align*}
Given  $p \in \partial \Sigma$ we want to find the  values $\sigma \in
\mathbb{C}$   and   tensors
$\widehat{\Psi}_{ij}$ at $p$ such that  $I_{\sigma} (\widehat{\Psi}) |_p$ is zero.
Assuming that  $I_{\sigma} (\widehat{\Psi}) |_p$ vanishes, a contraction of
$I_{\sigma} (\widehat{\Psi})_{ij} $ with $V^i$ gives
\begin{align*}
  (c_1 + 3) |V|^2_h V^i  \widehat{\Psi}_{ij}= 0.
\end{align*}
If $c_1 +3 \neq 0$ (i.e. $\sigma \not \in \{\frac{n-3}{2}, \frac{n-7}{2}\}$)
then necessarily $V^i \widehat{\Psi}_{ij} = 0$ and the condition
$I_{\sigma}(\widehat{\Psi}) =0$ reduces to $c_1   \widehat{\Psi}_{ij}=0$. If in addition $c_1 \neq 0$, the kernel is trivial. If $c_1=0$, i.e. $\sigma = \frac{n-1}{2}$ or
$\sigma= \frac{n-9}{2}$ the kernel is non-trivial and consists of tensor fields taking the form $\widehat{\Psi}_{ij} = X_i Y_j - X_j Y_i$ with covectors $X,Y$ annihilating $V$; equivalently, the associated vectors are tangent to $\partial \Sigma$.

If $c_1+ 3 =0$  then
$\widehat{\Psi}_{ij} := X_i V_j - X_j V_i$ (with $X$ as before) solves
$I_{\sigma} (\widehat{\Psi}_{ij})=0$:
\begin{align*}
  I_{\sigma} (\widehat{\Psi}_{ij}) & =
  I_{\sigma} (X_i V_j - X_j V_i ) \\
  & =
  c_1 |V|^2_h  (X_i V_j - X_j V_i )
   + 3 |V|^2_h X_i  V_j
   - 3 |V|^2_h X_j  V_i \\
   & = (c_1 +3 ) |V|^2_h
   \left ( X_i V_j - X_j V_i \right )  =0
    \,,
\end{align*}
which concludes the proof.
$\hfill \Box$
\bigskip

This leads to:

\begin{theorem}[Staticity extension, timelike region]
\label{t4XII21.1}
Assume that $(  \redM ,g)$ contains a bifurcate Killing horizon $\mcH$ and consider the
  quotient manifold $\Sigma$ in the region where $\xi$ is timelike.  Assume that the Ricci tensor of $(  \redM ,g)$
satisfies
$$
 d\big( g(\xi,\xi) \hat{r} \big) =0
$$
(in particular this is true for
$\Lambda$-vacuum spacetimes),
where the $1$-form $\hat r$ has been defined in \eqref{20III23.21}-\eqref{20III23.2}.
Then the tensor field
$\B_{ij}$ vanishes on $\partial \Sigma$ if and only if
the metric is static in the connected component adjacent to $\mcH$ of the region where the Killing vector is timelike; this is equivalent to  the  vanishing of $d\zeta$, where $\zeta$ is the torsion $1$-form of $S$.
\end{theorem}

\begin{Remark}
 \label{R22III23.1}
 {\rm
 It further follows from \eqref{24III23.85} that, for bifurcate horizons, this is equivalent to the vanishing of the transverse derivative of $\omega$ at any of the branches of the horizon.
\hfill $\Box$
}
\end{Remark}

{\noindent \sc Proof:}
Clearly $\B$ vanishes everywhere when the metric is static
    (cf.~\eqref{24III23.31}, \eqref{20III23.1} and \eqref{22III23.3}), in particular on $\partial \Sigma$. Thus the condition is necessary.

     To prove the converse implication, we start by noting that
for any $a\in\R$ the vanishing of $\B$ at
$\partial \Sigma$ is equivalent to $\C = o (u^a)$ at $\partial \Sigma$.

As already hinted-to, we choose $a=a_n\equiv (n-1)/2$.
The operator $L \equiv L_{(n-1)/2}$
is elliptic, uniformly degenerate and self-adjoint with respect
to the $L^2$ norm of the metric $\ov{h}$. For such an operator
 \cite[Theorem~A.14]{AILS} and \cite[Corollary (11)]{Mazzeo-continuation} apply. In particular \cite[Theorem~A.14 with $f\equiv 0$]{AILS} shows that
$\C$ is polyhomogeneous at $\partial \Sigma$.
  Matching coefficients in the polyhomogeneous expansion shows that $\B$ vanishes to infinite order at $\partial \Sigma$ because
$\C= o(u^a)$  and Lemma
\ref{indicial}.

The vanishing of $\C$ (and hence of $\B$) follows now from \cite[Corollary (11)]{Mazzeo-continuation}.

The relation to the vanishing of $d\zeta$ is the contents of Proposition~\ref{p22III22.2}.
\hfill $\Box$
\medskip

\subsection{The spacelike region}
 \label{ss24III23.1}

We continue our analysis in the region where $\epsilon=-1$, thus $u$ is a time function there. The aim is to find necessary and sufficient conditions for the twist to vanish in that region.

We consider a configuration  where the fields are smooth to the causal past of the bifurcation surface, or sufficiently differentiable as needed in the argument below, where finite differentiability suffices.  Such metrics can be constructed, for example, by solving  the  characteristic Cauchy problem with smooth KID data on a bifurcate null hypersurface, as described e.g.\ in \cite{ChPaetzKIDs,RaczHorizons2,KroenckePetersen}.

We formulate the staticity theorem in the past of a bifurcation surface, an identical result holds of course to the future:

\begin{theorem}[Staticity extension, spacelike region]
\label{T23XII21.1}
Let $\xi$ be a Killing vector defined on the past of an $(n-1)$-dimensional spacelike surface $S$ in an $(n+1)$-dimensional spacetime $(M,g)$, with $\xi=0$ on $S$.
  Assume that the Ricci tensor of $(  \redM ,g)$
satisfies
$$
 d\big( g(\xi,\xi) \hat{r} \big) =0
$$
(in particular this is true for
$\Lambda$-vacuum spacetimes).
Then
the Killing vector is static on the past domain of dependence $\mcD^-(S)$ if and only if the tensor field
$\B_{ij}$ vanishes on $\partial \Sigma$, if and only if the torsion covector of $S$ is closed.
\end{theorem}

{\noindent \sc Proof:}
The local calculations at $\partial \Sigma$ implicit in the proof of Theorem~\ref{t4XII21.1} show that $\C$ vanishes to all orders at $S$. In the analytic case  the result immediately follows.
In the smooth case one can instead rewrite the Fuchsian wave equation
  \eqref{mainEq}
as a symmetric hyperbolic system in a standard way, and use a large negative number $a\in\R$ in the definition of $\C$ so that the positivity condition of
 \cite[Lemma~4.14]{MPSS} holds.
\hfill$\Box$
\medskip

\section*{Acknowledgements} 

M.M. acknowledges financial support under
Grant PID2021-122938NB-I00 funded by MCIN/AEI /10.13039/501100011033 and
by “ERDF A way of making Europe” and RED2022-134301-T funded by
MCIN/AEI/10.13039/501100011033.

\appendix

\section{Appendix: Conformal behavior of the Laplacian on two-forms}

\label{ApA}
Consider an $n$-dimensional pseudo-Riemannian
manifold $(\Sigma,h)$. Let $u$ be a smooth positive function on $\Sigma$
and consider the metric 
$$
 \ov{h} := u^{-2} h
  \,.
$$
Objects defined with $\ov{h}$ carry an overline. Indices on such objects are moved with $\ov{h}$ while objects without an overline are manipulated using
$h$.

The following lemma relates $\Delta_h$ and
$\Delta_{\ov{h}}$ on two-forms.
\begin{lemma}
  \label{transConf}
  Let $\Psi$ be a smooth two-form on $\Sigma$. Then
  \begin{align}
    \Delta_{\ov{h}} \Psi_{ij}
    & = u^2 \Delta_h \Psi_{ij} + 2 \left ( u \Delta_h u + (2-n) |d u|^2_h
    \right ) \Psi_{ij} \nonumber \\
    & + (4-n) \left ( u D^k u \, D_k \Psi_{ij}
    + D^k u D_i u \, \Psi_{kj} - D^k u D_j u \, \Psi_{ki}
 \right ) \nonumber \\
    & -2 u \left ( D_i u (\delta \Psi)_j -
    D_j u (\delta \Psi)_i \right )
    + 2 u D^k u  (d \Psi)_{kij}. \label{ovDeltaDeltaPsi}
  \end{align}
\end{lemma}

Before going into the proof we need some preliminaries. We write
$Q$ for the difference tensor $Q(X,Y):= \ov{D}_X Y - D_XY$. Its explicit form is
\begin{align}
  Q^{k}{}_{il} &= \frac{1}{u} \left ( - \delta^k{}_i D_l u
- \delta^k{}_l D_i u + D^k u \, h_{il} \right ).
\label{Stensor}
\end{align}
Computing $\Delta_{\ov h}$ directly in terms of
$\Delta_h$ and the tensor $Q$ leads to long computations. We take a shortcut
via the Weizenb\"ock identity. It is convenient to
introduce the tensor
\begin{align}
H^k{}_{ilj} =-2 \Rsig^k{}_{ilj}+
\left ( \delta^k{}_l \Rsig_{ij}
- \delta^k{}_j \Rsig_{il}
+ \Rsig^k{}_l h_{ij}
  - \Rsig^k{}_j h_{il} \right )
  \label{defH}
\end{align}
and the corresponding tensor $\ov{H}^k{}_{ilj}$. The Weizenb\"ock
identity \eqref{Weiz} reads
\begin{align*}
\Delta_h \Psi_{ij} = - \Delta_{\mbox{\tiny dR}} \Psi_{ij} +
H^{k}{}_{i}{}^l{}_j  \Psi_{kl}
\end{align*}
so we can relate $\Delta_{\ov{h}}$ and $\Delta_h$ by
\begin{align}
  \Delta_{\ov{h}} \Psi_{ij} = u^2 \Delta_h \Psi_{ij}
  + \left ( \ov{H}^{k}{}_{i}{}^l{}_j
- u^2 H^{k}{}_{i}{}^l{}_j  \right )
  \Psi_{kl}
- \left ( \ov{\Delta}_{\mbox{\tiny dR}} \Psi_{ij} - u^2 \Delta_{\mbox{\tiny dR}} \Psi_{ij}
\right ).
  \label{relLaps}
\end{align}

 Let us momentarily assume   $n \geq 3$, we will consider the case   $n=2$ shortly.
In terms of the Schouten tensor $L$,
\begin{align*}
\Rsig_{ij} := (n-2) L_{ij} + L h_{jl}, \quad L := h^{ij} L_{ij}
 \,,
\end{align*}
and letting $\Csig^i{}_{ijk}$ be the Weyl tensor  of $h$, we have the decomposition
\begin{align*}
H^k{}_{ilj} = -2 \Csig^k{}_{ilj} -2
(4-n) \left ( \delta^k{}_l L_{ij}
- \delta^k{}_j L_{il}
+ L^k{}_l h_{ij}
- L h_{il} \right ) + 2 L \left ( \delta^k{}_l h_{ij}
- \delta^k{}_j h_{il} \right ).
\end{align*}
Under the conformal rescaling $\ov{h} = u^{-2}h$, the Schouten tensor transforms
as
\begin{align*}
\ov{L}_{ij} = L_{ij} + \frac{1}{u} D_{i}D_j u - \frac{|d u|^2_h}{2u^2}
h_{ij}.
\end{align*}
This, combined with conformal invariance of the Weyl tensor leads to
\begin{eqnarray}
 &
 \ov{H}^{k}{}_{i}{}^l{}_j
- u^2 H^{k}{}_{i}{}^l{}_j    =
h^{kl} W_{ij} - \delta^k{}_j W^l{}_i
+ W^{kl} h_{ij} - W^{k}{}_j \delta^l{}_i
 \,,
  &
    \label{relH}
\\
 &
 W_{ij}  := (n-4) u D_i D_j u + \left ( u \Delta_h u
+ (2-n) |d u|^2_h \right ) h_{ij}
 \,.
  &
   \label{expW}
\end{eqnarray}

Let us check that these equations also hold when $n=2$. Since the Riemann and Ricci tensors in two dimensions can be expressed in terms of the Gauss curvature and the metric, the tensor $H^k{}_{ilj}$ defined in \eqref{defH} is identically zero, and so is the difference left-hand side of  \eqref{relH}. Next, in dimension $n=2$ the tensor $W_{ij}$ defined in \eqref{expW} is trace-free. A tensor with the symmetries of a Riemann tensor in two-dimensions vanishes if and only if its double trace vanishes. The trace in
$\{ij\}$ and then $\{kl\}$  of the right-hand side of \eqref{relH} is
$2 \tr_h W$, hence zero. So, the right-hand side of \eqref{relH} is also zero and the validity of the expression in dimension $n=2$ is established.

In order to determine the conformal change of the de Rahm Laplacian we need to
relate the codifferentials $\ov{\delta}$ and $\delta$. Using the expression
$
(\delta T)_{i_2 \cdots i_p} = -D^{i_1} T_{i_1 \cdots i_p}$
on any $p$-form, an easy calculation leads to
\begin{align*}
\ov{\delta} T =
u^2  (\delta T)
+  (n-2p) i_{u \grad_h  u} T .
\end{align*}
So, for a two-form $\Psi$ we have
\begin{align*}
(d \circ \ov{\delta})  \Psi &= d ( u^2 (\delta \Psi)
+ (n -4) i_{u \grad_h u} \Psi )
= 2 u du \wedge (\delta \Psi) + u^2 (d \circ \delta) \Psi
+ (n-4) d \circ i_{u \grad_h u} \Psi  ,  \\
(\ov{\delta} \circ d) \Psi & = u^2 (\delta \circ d) \Psi
+ (n-6) i_{u \grad_h u} \circ d \Psi
 \,,
\end{align*}
and hence
\begin{align}
\ov{\Delta}_{\mbox{\tiny dR}} \Psi - u^2 \Delta_{\mbox{\tiny dR}} \Psi
=  2 du \wedge (\delta \Psi)
+  (n-4) \pounds_{u \grad_h u} \Psi
- 2 i_{u \grad_h u} \circ d \Psi
\,,
\label{ReldeRahm}
\end{align}
after using Cartan's identity $\pounds_V = d \circ i_V + i_V \circ d$.
With these identities at hand we may proceed with the proof of
Lemma \ref{transConf}.

\medskip

{\noindent \sc Proof:}
Insert \eqref{relH} and \eqref{ReldeRahm} into \eqref{relLaps} to get
\begin{align*}
\Delta_{\ov{h}} \Psi_{ij}
=&  u^2 \Delta_h \Psi_{ij}
+ W^k{}_i \Psi_{kj} - W^{k}{}_j \Psi_{ki}
- (n-4) \pounds_{u\grad_h u} \Psi_{ij} \\
&  -  2 u  \left ( D_{i} u (\delta \Psi)_{j}
- D_{j} u (\delta \Psi)_{i} \right )
+ 2 u D^k u  ( d\Psi)_{kij}.
\end{align*}
Expanding the Lie derivative and using the definition  \eqref{expW}
of $W$ yields \eqref{ovDeltaDeltaPsi}.
\hfill $\Box$
\bigskip

To continue, we write down two simple identities.
The first one is the
well-known conformal behaviour of the Laplacian on functions
 \begin{align}
   \Delta_{\ov{h}} f = u^2 \Delta_h f
   + (2-n) u D^i u D_i f
    \quad
   \Longrightarrow
   \quad
    \Delta_{\ov{h}} u =
   u^2 \Delta_h u + (2-n) u |d u|^2_h.
   \label{deltau}
 \end{align}
The second is an expression for $\ov{D}^l u \ov{D}_l \Psi_{ij}$. Using
$$
 D^l u \, Q^{k}{}_{il} = - u^{-1} |d u|^2_h \delta^k{}_i
$$
(which follows directly from \eqref{Stensor}) we get
 \begin{align}
   \ov{D}^l u \ov{D}_l \Psi_{ij} & = u^2 D^l u \left ( D_l \Psi_{ij}
   - Q^k{}_{il} \Psi_{kj} - Q^{k}{}_{jl} \Psi_{ik} \right ) \nonumber \\
   & = u^2 D^l u D_l \Psi_{ij}
   + 2 u |d u|^2_h \Psi_{ij}. \label{DuDPsi}
 \end{align}

The next lemma analyses self-adjointness of certain second order operators and
their  indicial maps. For this result we assume
    that $\Sigma$ is a manifold with boundary, that
    $h$ is positive definite, and that $u$ is a defining
    function for the boundary, i.e. $u=0$
    and $|d u|^2_h >0 $  on $\partial \Sigma$.
    \begin{lemma}
\label{indicialmap}
      With these assumptions, let $L (\Psi)_{ij} := \Delta_{\ov{h}} \Psi_{ij}
      + \ov{S}_{ij}{}^{kl} \Psi_{kl}$, with  $\ov{S}_{ij}{}^{kl}$ smooth up to and including the boundary. Suppose that
$\ov{S}_{ij}{}^{kl} = \ov{S}_{[ij]}{}^{kl} = \ov{S}_{ij}{}^{[kl]}$.
      Then
      \begin{itemize}
      \item[(i)] $L$ is formally self-adjoint (with respect to the $L^2$-scalar product
        with volume form defined by $\ov{h}$) if and only if
        \begin{align}
          \ov{S}\,^{kl}{}_{ij} = \ov{S}_{ij}{}^{kl}. \label{symmetry}
        \end{align}
      \item[(ii)] For $\sigma \in \mathbb{C}$, the indicial map $I_{\sigma} :
        \Lambda^2 (\Sigma) |_{\partial \Sigma} \otimes \mathbb{C} \rightarrow
          \Lambda^2 (\Sigma) |_{\partial \Sigma} \otimes \mathbb{C}$  is
            \begin{align}
              I_{\sigma} (\widehat{\Psi} )_{ij} = &
               \left (2(2-n) +\sigma ( 5 + \sigma -n) \right )
               |V|^2_h  \widehat{\Psi}_{ij}  \nonumber \\
& +  (4-n) \left ( V^k V_i  \widehat{\Psi}_{kj}
- V^k V_j \widehat{\Psi}_{ki} \right )
+ \ov{S}_{ij}{}^{kl} |_{\partial \Sigma}  \widehat{\Psi}_{kl}
\label{indicialExp}
            \end{align}
            where $V^i := D^i u |_{\partial \Sigma}$.
                  \end{itemize}
\end{lemma}

    {\noindent \sc Proof:}
Recall that the formal adjoint $L^{\dagger}$ is the (unique) operator that
for any pair of two-forms $\ov{\Psi}_{ij}$, $\ov{\Phi}_{ij}$  satisfies
(recall also that indices of objects with an overline are moved with $\ov{h}$)
\begin{align*}
\ov{\Psi}^{ij} L^{\dagger}(\ov{\Phi})_{ij}  -
\ov{\Phi}^{ij} L (\ov{\Psi})_{ij}  =
\ov{D}_k (\ov{F}^{k} (\ov{\Psi}, \ov{\Phi},
\ov{D} \ov{\Psi}, \ov{D} \Phi ))
\end{align*}
for some vector $\ov{F}^l$. It is immediate that
\begin{align*}
L^{\dagger} (\ov{\Phi})_{ij}
= \Delta_{\ov{h}} \ov{\Phi}_{ij} + \ov{S}^{kl}{}_{ij} \ov{\Phi}_{kl}
\end{align*}
because
\begin{align*}
\ov{\Psi}^{ij} L^{\dagger}(\ov{\Phi})_{ij}  -
\ov{\Phi}^{ij} L (\ov{\Psi})_{ij}  =
\ov{D}_k \left ( \ov{\Psi}^{ij} \ov{D}^k \ov{\Phi}_{ij}
- \ov{\Psi}^{ij} \ov{D}^k \ov{\Phi}_{ij} \right )
 \,.
\end{align*}

Consequently $L$ is formally self-adjoint ($L^{\dagger} = L$) if and only if \eqref{symmetry} holds. For item (ii) we compute $u^{-\sigma} \Delta_{\ov{h}} (u^{\sigma} \Psi)$ and get
    \begin{align*}
      \frac{1}{u^{\sigma}} \Delta_{\ov{h}} ( u^{\sigma} \Psi)_{ij}
      & = \left ( \frac{\sigma (\sigma-1)}{u^2} \ov{D}^ku \ov{D}_k u
      + \frac{\sigma}{u}  \Delta_{\ov{h}} u \right ) \Psi_{ij}
      + \frac{2 \sigma}{u} \ov{D}^{k} u \, \ov{D}_k \Psi_{ij}
      + \Delta_{\ov{h}} \Psi_{ij} \\
      & = \left ( \sigma u \Delta_h u
      + \sigma ( 5 + \sigma -n)
 |d u|^2_h \right ) \Psi_{ij}
      + 2 \sigma u D^k u \, D_k \Psi_{ij}
      + \Delta_{\ov{h}} \Psi_{ij}
       \,,
    \end{align*}
    where in the second equality we inserted  \eqref{deltau}, \eqref{DuDPsi} and $\ov{D}^l u \ov{D}_l u =
    u^2 |d u|^2_h$.
    By Lemma \ref{transConf} we have
    \begin{align*}
      \Delta_{\ov{h}} \Psi_{ij} |_{u=0} = \left .  2(2-n)
|d u|^2_h \Psi_{ij}
      + (4-n) \left ( D^k u D_i u \, \Psi_{kj}
- D^k u D_j u  \, \Psi_{ki}
\right )
\right |_{u=0}.
    \end{align*}
   The definition of indicial map gives \eqref{indicialExp}.

    $\hfill \Box$

The following corollary is used in the main text.
\begin{corollary}
\label{self}
Let $B_{ij}$ be a symmetric tensor and $c \in \mathbb{R}$. Then, the operator
\begin{align*}
L (\Psi)_{ij}
 = \Delta_{\ov{h}} \Psi_{ij}
+ c \Rsig^k{}_i{}^l{}_j \Psi_{kl} + B^k{}_i \Psi_{kj}
-  B^k{}_j \Psi_{ki}
\end{align*}
is formally self-adjoint with respect to the $L^2$ norm
of $\ov{h}$.
\end{corollary}
        {\noindent \sc Proof:}
One easily checks that the tensor
\begin{align*}
S_{ij}{}^{kl}
:=
c \Rsig^{[k}{}_{i}{}^{l]}{}_{j} - \delta^{[k}{}_j B^{l]}{}_{i}
   - B^{[k}{}_j \delta^{l]}{}_{i}
\end{align*}
is antisymmetric in $ij$ (the antisymmetry in $kl$ is obvious). In terms
of $S$ we can rewrite
$L (\Psi)_{ij} = \Delta_{\ov{h}} \Psi_{ij}
+ S_{ij}{}^{kl} \Psi_{kl}$
The symmetry condition \eqref{symmetry} can be checked with either
$\ov{h}$ or with $h$. We use $h$ and find
\begin{align*}
S^{ij}{}_{kl}
 =
c \Rsig_{[k}{}^i{}_{l]}{}^j  - \delta_{[k}{}^j B_{l]}{}^{i}
 - B_{[k}{}^j \delta_{l]}{}^{i}
=
c \Rsig^{[i}{}_k{}^{j]}{}_l - \delta^{[i}{}_l B^{j]}{}_{k}
 - B^{[i}{}_l \delta^{j]}{}_{k}  = S_{kl}{}^{ij}
\end{align*}
after using the symmetries of the Riemann tensor and of $B$.
    $\hfill \Box$
\bigskip

\bibliographystyle{amsplain}
\bibliography{ChruscielMarsStaticity-minimal}

\end{document}